# Astrodynamical Space Test of Relativity using Optical Devices I (ASTROD I) - A class-M fundamental physics mission proposal for Cosmic Vision 2015-2025: 2010 Update

Claus Braxmaier
*IOS Institute for Optical Systems, Faculty of Mechanical Engineering, HTWG Konstanz, University of Applied Sciences, Konstanz, Germany*

Hansjörg Dittus
*Institute of Space Systems, German Aerospace Center, Robert-Hooke-Strasse 7, 28359 Bremen, Germany*

Bernard Foulon
*Office National D'Édudes et de Recherches Aerospatiales (ONERA), BP 72 F-92322 Chatillon Cedex, France*

Ertan Göklü
*Center of Applied Space Technology and Microgravity (ZARM), University of Bremen, Am Fallturm, 28359 Bremen, Germany*

Catia Grimani
*DiSBeF Department, Universtità degli Studi di Urbino 'Carlo Bo', Urbino (PU) and INFN Florence Italy*

Jian Guo
*Delft University of Technology (TU Delft), The Netherlands*

Sven Herrmann
*Center of Applied Space Technology and Microgravity (ZARM), University of Bremen, Am Fallturm, 28359 Bremen, Germany*

Claus Lämmerzahl
*Center of Applied Space Technology and Microgravity (ZARM), University of Bremen, Am Fallturm, 28359 Bremen, Germany*
e-mail: claus.laemmerzahl@zarm.uni-bremen.de

Wei-Tou Ni
*Shanghai United Center for Astrophysics, Shanghai Normal University, No. 100, Guilin Road, Shanghai, 200234, China*
e-mail: weitou@gmail.com

Achim Peters
*Department of Physics, Humboldt-University Berlin, 10117 Berlin, Germany*

Benny Rievers
*Center of Applied Space Technology and Microgravity (ZARM), University of Bremen, Am Fallturm, 28359 Bremen, Germany*

Étienne Samain
*Observatoire de la Côte d'Azur, UMR Gemini, R&D Métrologie, 06460 Caussols, France*

Hanns Selig
*Center of Applied Space Technology and Microgravity (ZARM), University of Bremen, Am Fallturm, 28359 Bremen, Germany*

Diana Shaul
*High Energy Physics Group, Blackett Laboratory, Imperial College London, Prince Consort Road, London, SW7 2BZ, UK*

Drazen Svehla
*ESA/ESOC, Navigation Office, Darmstadt, Germany*

Pierre Touboul
*Office National D'Édudes et de Recherches Aerospatiales (ONERA), BP 72 F-92322 Chatillon Cedex, France*

Gang Wang
*Purple Mountain Observatory, Chinese Academy of Sciences, No. 2, Beijing W. Road, Nanjing, 210008, China*

An-Ming Wu
*National Space Organization (NSPO), 8F, 9 Prosperity 1st Road, Science Park Hsinchu, Taiwan, 30078, ROC*

Alexander F. Zakharov
*Institute of Theoretical and Experimental Physics, 25, B. Cheremushkinskaya St., Moscow, 117259, Russia*

**Abstract:** This paper on ASTROD I is based on our 2010 proposal submitted for the ESA call for class-M mission proposals, and is a sequel and an update to our previous paper [Experimental Astronomy 23 (2009) 491-527; designated as Paper I] which was based on our last proposal submitted for the 2007 ESA call. In this paper, we present our orbit selection with one Venus swing-by together with orbit simulation. In Paper I, our orbit choice is with two Venus swing-bys. The present choice takes shorter time (about 250 days) to reach the opposite side of the Sun. We also present a preliminary design of the optical bench, and elaborate on the solar physics goals with the radiation monitor payload. We discuss telescope size, trade-offs of drag-free sensitivities, thermal issues and present an outlook.

ASTROD I is a planned interplanetary space mission with multiple goals. The primary aims are: to test General Relativity with an improvement in sensitivity of over 3 orders of magnitude, improving our understanding of gravity and aiding the development of a new quantum gravity theory; to measure key solar system parameters with increased accuracy, advancing solar physics and our knowledge of the solar system; and to measure the time rate of change of the gravitational constant with an order of magnitude improvement and the anomalous Pioneer acceleration, thereby probing dark matter and dark energy gravitationally. It is envisaged as the first in a series of ASTROD missions. ASTROD I will consist of one spacecraft carrying a telescope, four lasers, two event timers and a clock. Two-way, two-wavelength laser pulse ranging will be used between the spacecraft in a solar orbit and deep space laser stations on Earth, to achieve the ASTROD I goals.

For this mission, accurate pulse timing with an ultra-stable clock, and a drag-free spacecraft with reliable inertial sensor are required. T2L2 has demonstrated the required accurate pulse timing; rubidium clock on board Galileo has mostly demonstrated the required clock stability; the accelerometer on board GOCE has paved the way for achieving the reliable inertial sensor; the demonstration of LISA Pathfinder will provide an excellent platform for the implementation of the ASTROD I drag-free spacecraft. These European activities comprise the pillars for building up the mission and make the technologies needed ready.

A second mission, ASTROD or ASTROD-GW (depending on the results of ASTROD I), is envisaged as a three-spacecraft mission which, in the case of ASTROD, would test General Relativity to one part per billion, enable detection of solar g-modes, measure the solar Lense-Thirring effect to 10 parts per million, and probe gravitational waves at frequencies below the LISA bandwidth, or in the case of ASTROD-GW, would be dedicated to probe gravitational waves at frequencies below the LISA bandwidth to 100 nHz and to detect solar g-mode oscillations. In the third phase (Super-ASTROD), larger orbits could be implemented to map the outer solar system and to probe primordial gravitational-waves at frequencies below the ASTROD bandwidth.

**Keywords** Probing the fundamental laws of spacetime • Exploring the microscopic origin of gravity • Testing relativistic gravity • Mapping solar-system gravity • Solar g-mode detection • Gravitational-wave detection • ASTROD • ASTROD I • ASTROD-GW

**PACS** 04.80.Cc • 04.80.Nn • 95.10.-a • 96.60.Ly

## 1. Introduction and mission goals

This paper is a follow-up and update of our previous paper [1] (designated as Paper I) on ASTROD I. The general concept of ASTROD (Astrodynamical Space Test of Relativity using Optical Devices) is to have a constellation of drag-free spacecraft (S/C) navigate through the solar system and range with one another and/or ground stations using optical devices to map and monitor the solar-system gravitational field, to measure related solar-system parameters, to test relativistic gravity, to observe solar g-mode oscillations, and to detect gravitational waves. The gravitational field in the solar system influences the ranges and is determined by three factors: the dynamic distribution of matter in the solar system; the dynamic distribution of matter outside the solar system (galactic, cosmological, etc.); and gravitational waves propagating through the solar system. Different relativistic theories of gravity make different predictions of the solar-system gravitational field. Hence, precise measurements of the solar-system gravitational field test these relativistic theories, in addition to gravitational wave observations, determination of the matter distribution in the solar-system and determination of the observable (testable) influence of our galaxy and cosmos. Since ranges are affected by gravitational fields, from the precise range observations and their fitting solution, the gravitational field in the solar system with different contributing factors can be determined with high precision. Two keys in this mission concept are range measurement accuracy and minimizing spurious non-geodesic accelerations. The range accuracy depends on timing uncertainty, while minimizing spurious non-geodesic accelerations relies on drag-free performance.

A baseline implementation of ASTROD (also called ASTROD II) is three-spacecraft mission with two spacecraft in separate solar orbits and one spacecraft near Earth (Sun-Earth L1/L2 point), each carrying a payload of a proof mass, two telescopes, two 1–2 W lasers, a clock and a drag-free system and range coherently with one another using lasers [2, 3]. ASTROD I using one spacecraft ranging with ground laser stations is the first step toward the baseline mission ASTROD [1, 4, 5]. In the Cosmic Vision version of ASTROD I mission concept [1], we use 2-color laser pulse ranging and assume a timing uncertainty of 3 ps together with residual acceleration noise requirement for 2 orbit plane axes

$$S_{\Delta a}^{1/2}(f) = 1 \times 10^{-14} [(0.3 \text{ mHz}/f + 30 \times (f/3 \text{ mHz})^2] \text{ m s}^{-2} \text{ Hz}^{-1/2}, \qquad (1)$$

over the frequency range of 0.1 mHz < $f$ < 100 mHz with $f$ in unit of mHz. For the perpendicular axis, this condition can be relaxed by one order of magnitude. The 3 ps timing accuracy is already achieved by T2L2 (Time Transfer by Laser Link) event timer [6, 7] on board Jason 2 [8]. The 2-color laser ranging is to measure and subtract atmospheric refraction delay. The range measurement accuracy is 0.9 mm (3 ps) and has been verified in SLR (Satellite Laser Ranging) and LLR (Lunar Laser Ranging) [9, 16]; compared with the maximum S/C distance to Earth of 1.7 AU, the fractional accuracy is $0.3 \times 10^{-14}$ in range. This is more than 3 orders of magnitude improvement over radio range experiment which has an accuracy of a couple of meters at present. Test of relativistic gravity and measurement of solar system parameters will be improved by more than 3 orders of magnitude. The corresponding drag-free performance (1) is required in company to achieve this improvement.

ASTROD I mission concept has one spacecraft carrying a payload of a telescope, four lasers (two spares), and a clock ranging with ground stations (ODSN: Optical Deep Space Network). A schematic of the payload configuration of ASTROD I is shown in Fig. 1 of Paper I. With a cesium/rubidium clock on board spacecraft, a fractional precision of distance determination of $10^{-14}$ will be achieved. A mission summary is compiled in Table 1 of Paper I.

The scientific aims include a better measurement of the relativistic parameters and many solar system parameters. For a launch in 2021, after one encounter with Venus, the orbital period can be shortened to 182.3 days [9]. The science measurement starts immediately after orbit correction [1, 5]. The range measurement can be performed a couple of times a day or one time in a couple of days. Since the measurement principle does not depend on cancellation, some deviation from the fiducial orbit design is allowed. The Venus flyby is for shortening the time to reach the other side of the Sun (and to measure the Venus gravity field). After about 250 days from launch, the spacecraft will arrive at the other side of the Sun for a precise determination of the light deflection parameter γ through Shapiro time delay measurement so that the correlation of this parameter and the relativistic nonlinear parameter β can be decreased. The configuration after about 250 days from launch is sketched in Figure 1.

More specifically, *the scientific goals of ASTROD I are threefold.* The first goal is to test relativistic gravity and the fundamental laws of spacetime with an improvement of more than three orders of magnitude in sensitivity, specifically, to measure the PPN (Parametrized Post-Newtonian) parameter γ (Eddington light-deflection parameter; for general relativity, it is 1) via Shapiro time delay to $3 \times 10^{-8}$, β (relativistic nonlinear-gravity parameter; for general relativity, it is 1) to $6 \times 10^{-6}$ and others with significant improvement; to measure the fractional time rate of change of the gravitational constant (dG/dt)/G with two orders of magnitude improvement; and to measure deviations from the Einsteinian gravitational acceleration, i.e., an anomalous Pioneer acceleration $A_a$ [20], with several orders of magnitude improvement. The second goal is to initiate a revolution of astrodynamics with laser ranging in the solar system. The third goal is to increase the sensitivity of solar, planetary and asteroid parameter determination by 1 to 3 orders of magnitude. In this context, the measurement of solar quadrupole moment parameter $J_2$ will be improved by two orders of magnitude, i.e., to $10^{-9}$. Table 1 gives a summary of these objectives. The present accuracies of the parameters are listed in the second column of the table.

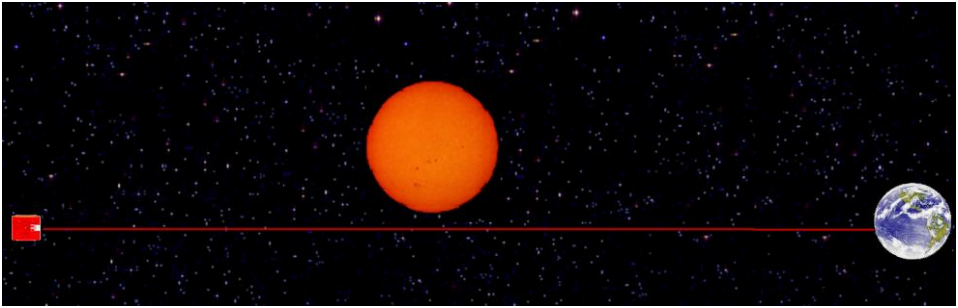

**Fig. 1** A sketch of ASTROD I mission at about 250 days after launch (Sun located near the line between ASTROD I and Earth). The spacecraft is at about 1.7 AU from Earth at this time.

**Table 1** Summary of the scientific objectives of the ASTROD I mission

| Effect/Quantity | Present accuracy [10] | Projected accuracy |
|---|---|---|
| PPN parameter β | $2 \times 10^{-4}$ | $6 \times 10^{-6}$ |
| PPN parameter γ (Eddington parameter) | $4.4 \times 10^{-5}$ | $3 \times 10^{-8}$ |
| (dG/dt)/G | $10^{-12}$ yr$^{-1}$ | $1 \times 10^{-14}$ yr$^{-1}$ |
| Anomalous Pioneer acceleration $A_a$ | $(8.74 \pm 1.33) \times 10^{-10}$ m/s$^2$ | $6 \times 10^{-16}$ m/s$^2$ |
| Determination of solar quadrupole moment parameter $J_2$ | $(1.82 \pm 0.47) \times 10^{-7}$ [11] | $1 \times 10^{-9}$ |
| Measuring solar angular momentum via solar Lense-Thirring Effect | 0.1 | 0.1 |
| Determination of planetary masses and orbit parameters | (depends on object) | 1 - 3 orders better |
| Determination of asteroid masses and density | (depends on object) | 2 - 3 orders better |

These goals are based on the orbit simulation for an orbit suitable for launch in 2021 with timing uncertainty 3 ps, and drag-free uncertainty is $3 \times 10^{-14}$ m s$^{-2}$ Hz$^{-1/2}$ at 0.1 mHz [14]. The simulation is from 130 days after launch for 1000 days with 5 observations per day. In the numerical integration of for this orbit simulation, we have taken the following effects into account:
(i) Newtonian and post-Newtonian point mass gravitational interaction between every two bodies including the Sun, nine planets, Moon, three big asteroids (Ceres, Pallas and Vesta), with additional 349 asteroids included as sources non-interactively
(ii) Newtonian attraction between a body with gravitational multipoles including Sun ($J_2$), Earth ($J_2$, $J_3$, $J_4$) and Venus ($J_2$), and others as point masses
(iii) the spacecraft is treated as a test body in the numerical integration.

The details of this orbit choice are presented in section 2. Other similar choices of orbit have similar results.

The results of simulations in section2 are shown in Table 1. When the observation times in the orbit simulation is decreased to one time per day instead of five times per day, the uncertainty increased by about 10 %. A more thorough orbit simulation including the Venus flyby and the period after injection orbit correction is underway.

In addition to the above primary scientific objectives, *important goals can be achieved on solar physics using radiation monitors needed for charge management on board the ASTROD missions* (section4).

For ASTROD I, the critical technologies are (i) drag-free control; (ii) interplanetary pulse laser ranging and (iii) sunlight shielding. The drag-free requirement for ASTROD I is relaxed by one order of magnitude as compared with that of LISA (Laser Interferometer Space Antenna for gravitational-wave detection) [12]. The drag-free technologies under development for LISA Pathfinder [13] will largely meet the requirement of ASTROD I. The successful functioning of the accelerometer on board GOCE [14] reassured the development for achieving improved inertial sensors. Since the solar array is fixed (non-moving) around the cylindrical side of the spacecraft [1], it will not cause problems with the drag free system. Millimetre precision has already been achieved in lunar laser ranging [15, 16]. Interplanetary pulse laser ranging (both up and down) was demonstrated by MESSENGER, using its laser altimeter in 2005 [17]. The technologies needed for a dedicated mission using interplanetary pulse laser ranging with millimetre accuracy are already mature. Sunlight shielding is a common technology which needs to be developed for optical missions measuring Shapiro time delay and light deflection due to the Sun. The technological readiness of ASTROD I is relatively high and it could fit into ESA's proposed Drag-Free Fundamental Physics Explorer Mission Series.

*In section 2, we present our orbit selection with one Venus swing-by together with orbit simulation. In Paper I, our orbit choice is with two Venus swing-bys. The present choice takes shorter time (about 250 days) to reach the opposite side of the Sun. In section 3, we present a preliminary design of optical bench. In section 4, we elaborate on the solar physics goals with the radiation monitor payload. In section 5, we make a number of discussions and give an outlook.*

**2. Orbit selection and orbit simulation**

The spacecraft needs to reach the far side of the Sun to measure the Shapiro effect (relativistic light retardation) to high precision. A trade-off is needed between minimising the time to reach the far side of the Sun because of the risk to the mission from aging of equipment, and reducing the amount of fuel needed for the mission. If the spacecraft is launched into solar orbit with a period about 300 days at a suitable epoch, the perihelion can reach 0.72 AU to encounter Venus. At this point we have two choices: (i) A deep encounter to swing the spacecraft to a perihelion of about 0.5 AU or deeper; (ii) If we tune the Venus encounter such that the spacecraft changed into a 225-day orbit, the spacecraft has a chance to have a second encounter with the Venus one-period or half a period later with a good swing toward the Sun. Our design concept saves fuel and the spacecraft still reaches the other side of the Sun quickly. The science observation starts within a couple of months from launch after the orbit injection and orbit corretion are done. Drag-free operation will be operative through the whole science observation. In our Cosmic Vision proposal last time, we take a choice from (ii). Here we present a choice from (i) in the following [9].

Earth's orbit (1 AU) around the Sun has a period of 365 days. Venus' orbit (semimajor axis 0.723 AU) around the Sun has a period of 225 days. The synodic period of Venus is 584 days. For a launch of a spacecraft into a solar orbit with aphelion around 1 AU (Earth) and perihelion around 0.723 AU (Venus), the semi-major axis is 0.808 AU. By Kepler's law, the period of this orbit is 294 days and it takes around 147 days to reach the Venus. At the launch position, Earth should be ahead of Venus by 55° so that Venus could catch the spacecraft just in time. Around October, 2021, Earth and Venus are in such relative positions and launches are possible. With these simple calculations, the Venus encounter would be around 150 days after launch. Because of the eccentricity and inclination of Earth and Venus orbits, and the small but nonzero angles of encounter, actual numbers have a range, and we can take the advantage of this to minimize the time for reaching the position of encounter. However, these numbers serve as starting points for orbit design. After some efforts, we found the following choice of initial state given below in solar-system barycentric equatorial coordinate system:

Initial epoch: 2021-Oct-15 08:49:55.200 (JD 2,459,502.868)
Initial position: X = 0.916,698,401,927,073,3 AU
 Y = 0.345,242,519,061,967,7 AU
 Z = 0.149,844,674,914,063,6 AU
Initial velocity: $V_X$= -0.007,798,911,029,395,402 AU/day
 $V_Y$= 0.010,460,445,237,373,31 AU/day
 $V_Z$= 0.011,332,904,469,003,998 AU/day

With this initial state at about 500 km height from Earth surface, the orbit obtained is shown in Fig. 2 - Fig. 5. Fig. 2 shows the orbit in the X-Y plane of the heliocentric ecliptic coordinate system. The distance between spacecraft and Venus as a function of mission day is shown in Fig. 4. The Venus encounter is at 112.02 days from starting epoch and with a distance to the Venus center of 10920 km. The apparent position of spacecraft reaches the opposite side of the Sun after 250.0 days, 613.2 days and 976.3 days from launch, and the orbit in the Sun-Earth fixed frame is shown as Fig. 5. The apparent angles of the spacecraft during the three solar oppositions are shown in Fig. 6.

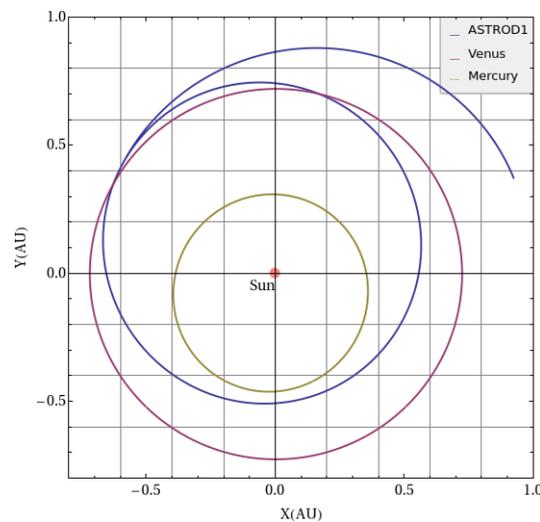
Fig. 2. The 2021 orbit in heliocentric ecliptic coordinate system

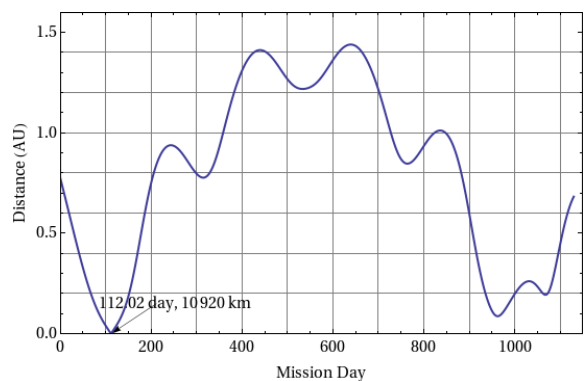
Fig. 3. Distance between spacecraft and Venus

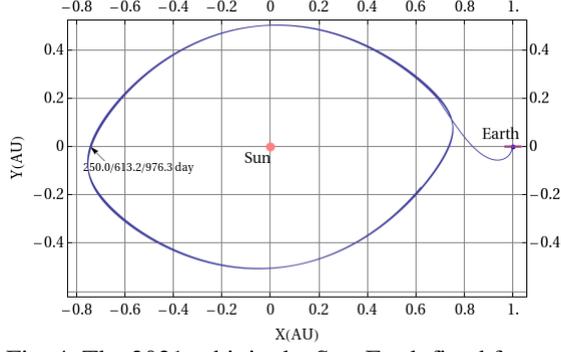

Fig. 4. The 2021 orbit in the Sun-Earth fixed frame.

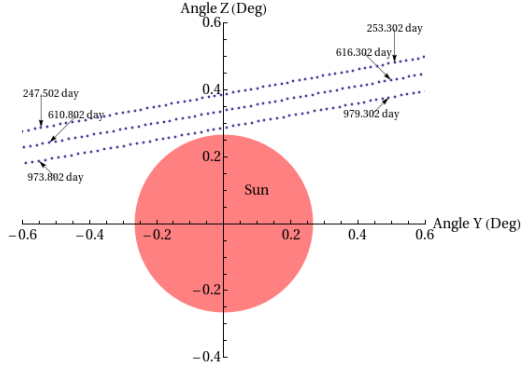

Fig. 5. Apparent angles during the three solar oppositions

The one-way Shapiro time delays are shown in Fig. 6. Near the three solar oppositions, the maximum Shapiro time delays are 0.1091 ms, 0.1118 ms and 0.1149 ms respectively.

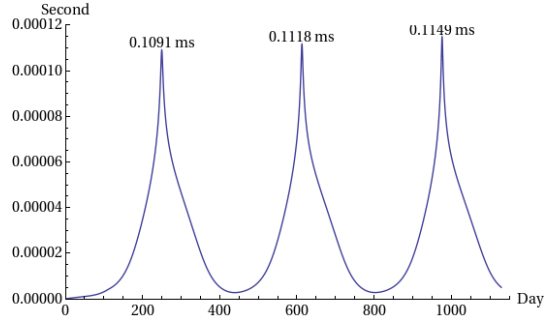

Fig. 6. Shapiro time delays

We use the following stochastic model to simulate the ranging time between ASTROD I spacecraft and Earth. We consider two kinds of noises. The first kind is the imprecision of optical ranging devices. It will influence ranging time directly. This part is treated as a Gaussian random noise with zero mean and with magnitude $3\times10^{-12}$ sec. The second kind is unknown accelerations due to imperfection of the spacecraft drag free system. The magnitude of unknown acceleration is treated as a Gaussion random noise with zero mean and with half width $3\times10^{-16}$ m/s$^2$; this is more or less equivalent to a noise spectrum density of $3 \times 10^{-14}$ m s$^{-2}$ Hz$^{-1/2}$ at 0.1 mHz. We change the direction of unknown acceleration randomly every four hours [2, 9].

We simulated the orbit from 130 days after launch (18 days after Venus encounter) to 1130 days after launch for 1000 days after the spacecraft arriving at its destination orbit. Adding the two kind of noise into the orbit calculation program, we obtain simulated data set; in Fig. 7, we show 2 examples of deviations of simulated range from the fiducial range. The parabolic error curves in Fig. 7 are valid for short segments of trajectories, and represent over-estimation for periodic/quasi-periodic motions. This issue together with relaxation of acceleration noise requirement on the axis perpendicular to the orbit plane is dealt with in [9]. The uncertainties quoted in this section are over-estimations for longer times.

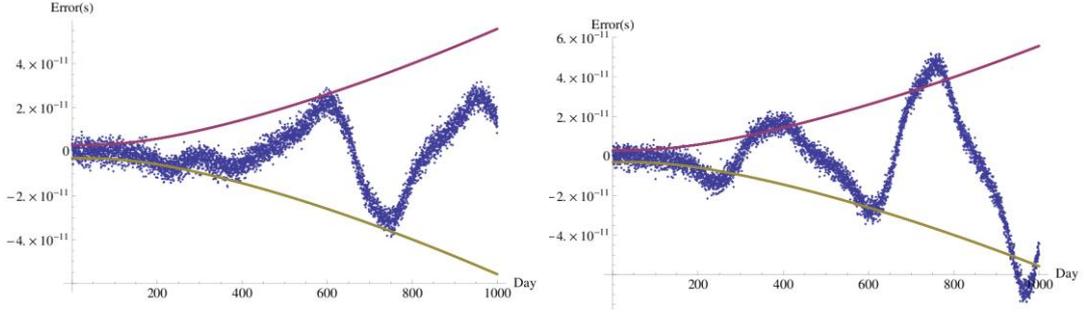

Fig. 7. Two examples of deviations of simulated range from the fiducial range

We estimated the uncertainties of 25 parameters together with parameter correlations as a function of epochs using sequencial Kalman filter data processing. The uncertainty of γ is shown in Fig. 8. The correlation between γ and β as a function of epochs is shown in Fig. 9. The uncertainty of γ has a steep decent during the first solar opposition. Table 2 shows the uncertainty of 25 parameters obtained from Kalman algorithm. The uncertainty for β is different from [1] for in the calculation quoted there is an error in the partials for β

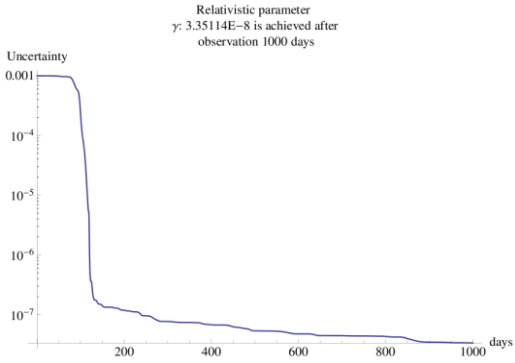 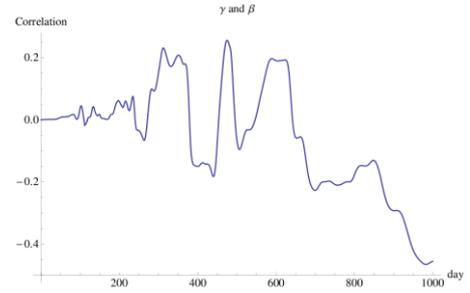

Fig. 8. Uncertainty of γ as a function of epoch.   Fig. 9. Correlation between γ and β

Now we explain in more detail how the Lense-Thirring effect induced from the time delay is to be measured and in particular how the effects induced by gravitational mass and quadrupole moments are discriminated. The Lense-Thirring effect on the propagation time of the pulse signals changes sign from one side of Sun to the other side of the Sun; it also changes sign for signals to the spacecraft and from the spacecraft. However, the effects on the propagation time due to mass and quadrupole moments do not change sign. We have done a simulation including Lense-Thirring effect, and verify the estimate that it can be determined to 10 % or somewhat better [9]. The correlation with effects due to gravitational mass and quadrupole moments are very small. The degradation (increase) in the uncertainty of determining the Eddington parameter γ is less than 2 % with and without fitting the Lense-Thirring effect. Since the Lense-thirring effect due to Earth is already verified to 10 % or better, this gives a solar angular momentum measurement to 10 % (or better) which is useful as a consistency check with solar modelling value based on solar seismology.

The Soyuz-Fregat launch vehicle has the performance to lift 1100 kg to GTO. CZ-2C or CZ-4B, has the performance to lift 1000 kg or 1500 kg to GTO, respectively. These launchers could be considered as candidate launchers.

In the orbit design of the ASTROD I spacecraft for the 2021 launch windows, we achieved our goal of design to have gravity-assistance once from Venus to reach shorter period for making the relativistic Shapiro time sooner. From the result of 2021 orbit simulation, the goal of ASTROD I for improvement in determining relativistic parameters and measuring solar-system parameters can be achieved.

Tabel 2. The Uncertainties of Parameters

| Parameter | | Initial value of parameter | Uncertainty |
|---|---|---|---|
| $\gamma$ | | 1.0 | 3.35028E-08 |
| $\beta$ | | 1.0 | 6.62060E-06 |
| $J_2$ of Sun | | 2.0E-7 | 1.03118E-09 |
| Gm of celestial body ($AU^3/day^2$) | Mercury | 4.912547451450812E-11 | 1.32278E-19 |
| | Venus | 7.243452486162703E-10 | 1.02028E-20 |
| | Earth | 8.887692390113509E-10 | 2.71371E-22 |
| | Mars | 9.549535105779258E-11 | 2.49856E-19 |
| | Jupiter | 2.825345909524226E-7 | 1.24532E-16 |
| | Saturn | 8.459715185680659E-8 | 1.86588E-15 |
| | Uranus | 1.292024916781969E-8 | 8.89023E-14 |
| | Neptune | 1.524358900784276E-8 | 1.97430E-13 |
| | Pluto | 2.188699765425970E-12 | 4.17084E-13 |
| | Moon | 1.093189565989898E-11 | 9.96041E-23 |
| | Sun | 2.959122082855911E-4 | 4.48670E-17 |
| | Ceres | 1.390787378942278E-13 | 5.06383E-18 |
| | Pallas | 2.959122082855911E-14 | 4.68253E-18 |
| | Vesta | 3.846858707712684E-14 | 2.64966E-18 |
| Initial state of spacecraft | X (AU) | -0.662821258577325564 | 1.16421E-14 |
| | Y (AU) | -0.015742050207228120 | 2.77439E-14 |
| | Z (AU) | 0.0338609441654262801 | 9.83737E-14 |
| | Vx (AU/day) | 0.0042454105224505364 | 3.90780E-16 |
| | Vy (AU/day) | -0.018474310473282873 | 5.40500E-16 |
| | Vz (AU/day) | -0.0086066308819272103 | 1.41347E-15 |
| (dG/dt)/G (yr$^{-1}$) | | 0.0 | 7.46304E-16 |
| $A_a$ (m/s$^2$) | | 0.0 | 5.75425E-16 |

## 3. Optical Bench

A preliminary design of the optical bench (OB) is shown in the following Fig. 10. It is based on the schematic as shown in Fig. 6 of Paper I. The OB is foreseen to be mounted vertically behind the telescope; the inertial sensor is mounted to the backside of the OB. Therefore, the laser beams to and from the telescope are deflected by 90° into the plane of the OB by a mirror. The laser beam on the OB is a collimated beam with a beam diameter of 1 to 2 mm.

Both laser wavelengths (1064nm and 532nm) are realized in redundancy, two lasers for each frequency are out coupled on the OB. A fiber switch is implemented for switching between the two redundant laser sources. Its design was developed in the LISA context: The beams from the two lasers – which are orthogonally polarized to each other – are superimposed at a polarizing beamsplitter (PBS). When changing the laser source, the wave plate must be rotated by 45 degrees. A polarizer guarantees the clean and proper polarization on the OB. A design of a rotatable waveplate is developed in the LISA context on EM level. The detectors are realized in hot redundancy, the laser beams are split in front of the avalanche photo detectors; analysis shows the feasibility especially in terms of optical power. Dichroitic beam splitters are used for multiplexing the two laser frequencies. Additional photo detectors are implemented after fiber out-coupling of the laser sources for phase and timing reference.

The material of the breadboard and the corresponding assembly-integration technology is TBD; options are ULE, Zerodur, Titanium, Invar, Auminum, SiC, ... Compared to e.g. the LISA OB, where pm- and nrad-stability is required, ASTROD I has much lower requirements in translation measurement. Allocating one tenth of the 0.9mm range accuracy to the OB, this corresponds to 90µm stability in optical pathlength on the OB. Assuming a T = ± 1K temperature stability of the OB and an optical pathlength of 20cm, a very moderate CTE $< 2.2*10^{-4}$ is required for the OB material and the AI-technology. While also aluminum can deliver this stability, specific AI technologies have been developed for LISA Pathfinder, LISA,... using a baseplate made of Zerodur where the optical components are fixed by using hydroxide-catalysis bonding technology. This technology is proven and space-qualified.

The given mass of 5kg is a rough estimate, based on the size and complexity of the optical bench. Different materials (aluminum, Zerodur, ...) will result in similar values.

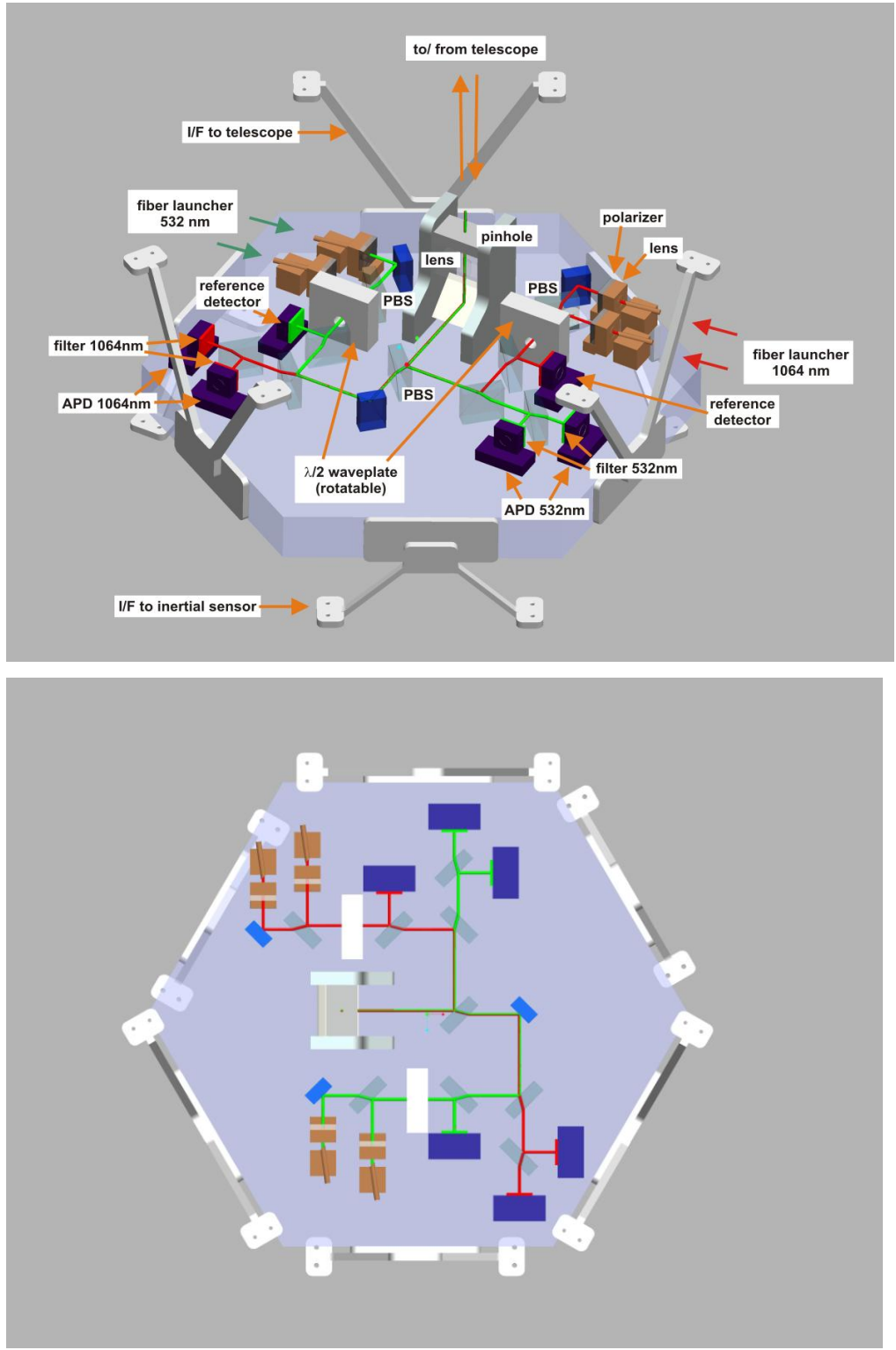

Fig. 10. Optical bench design for ASTROD I.

## 4. Solar physics using radiation monitors

Radiation detectors are needed to monitor the test-mass charging on board ASTROD missions. In this section, we present the important goals that can be achieved on solar physics, space weather and related disciplines using these radiation monitors.

Galactic and solar energetic particles (SEPs) pose a hazard to manned and unmanned missions. Galactic cosmic rays (GCRs) consist basically of 90% of protons, 8% of helium nuclei, 1% of heavy nuclei and 1% electrons. The energy range varies between 100 MeV(/n) and above $10^{20}$ eV. The energy spectra of solar particles, showing an analogous composition, are much softer and only the strong events are characterized by particle accelerated at GeV energies. Both GCRs and solar particles show latitude, longitude and distance dependence from the Sun. The magnitude of galactic cosmic ray gradients varies from 10 % per AU at 1 AU to 4 % per AU at 5 AU.

An accurate study of cosmic ray variations and fluctuations and solar event occurrence is mandatory to evaluate the performance of missions like LISA and ASTROD. Very often radiation monitors are placed on board space missions in order to measure the galactic and solar particle fluxes reaching the spacecraft at each time.

This is the case of LISA and its precursor mission LISA Pathfinder. The silicon detectors that will be placed on LISA Pathfinder [18] will allow for proton and helium nucleus detection above a few tens of MeV(/n). Ions with energies larger than 100 MeV(/n) penetrate the spacecraft and, therefore, particle detector observations will help to estimate the proof mass charging [19, 20]. In particular, LISA simulations indicate that solar events with fluences larger than $10^7$ protons/cm$^2$ overcome the whole mission noise budget in the low frequency range in a few tens of minutes ([21] and references therein).

Galactic cosmic-ray energy spectra and occurrence of solar events at the time of future space missions can be estimated on the basis of the projections of the coming solar cycles. Due to the long duration of the last solar minimum, for example, the solar cycle 24 is expected to be quite weak ([22] and references therein). As an example, the number of solar events with fluences larger than $10^6$ protons/cm$^2$ expected in six months of the LISA Pathfinder data taking during the second half of 2012 ranges between 1.5 and 3.4 [22]. Each event lasts a few days. It was evaluated if it was worthwhile to modify the LISA-PF radiation monitors to add SEP forecasting capabilities to LISA [21]. For the ASTROD I mission the test mass charging studies were reported in [23-24] for both galactic cosmic-ray and solar energetic particles. In ASTROD I the test masses are surrounded by approximately the same amount of matter of the LISA missions and therefore similar results were obtained for the charging. Analogously, strong solar events were found to saturate the whole mission noise budget.

In addition to the main role of noise monitoring, particle detectors placed on board missions for gravitational wave detection in space would provide precious clues on solar physics for space weather applications. We recall that space weather investigations have fundamental importance for solar event forecasting affecting air transportation, life in space, communications, etc.

In particular, cosmic rays and solar particles with energies larger than 100 MeV/n penetrate spacecraft and astronaut suits. The August 1972 flare would have compromised the health of a crew on the surface of the Moon. Similar exposure to giant flare, such as that recorded in September 1859 might even result in death. Unfortunately, because of the difficulty to separate galactic cosmic rays from solar particles above 100 MeV(/n) and since only the stronger solar events generate particles at these energies, the majority of experiments intended for solar physics do not allow for high energy particle discrimination from lower energy ones. Consequently, very few data are available in the literature on SEP high energy differential fluxes. Recently, a satellite experiment for cosmic-ray antimatter detection, PAMELA [25], measured the proton and helium differential fluxes associated with the solar event dated December 13th 2006. However, PAMELA is in flight since June 15th 2006 during the very long lasting solar minimum and therefore, did not detect many solar events. This mission will end before the next solar maximum. AMS [26] is the next experiment expected to carry a magnetic spectrometer in space for differential cosmic-ray flux measurements.

LISA and ASTROD are natural multipoint observatories. LISA Pathfinder will observe SEP variations at $1.5 \times 10^6$ km from Earth. The LISA three spacecraft constellation will cover 2 degrees in longitude at 1 AU and 20 degrees in longitude will separate LISA from Earth. The expected duration of 10 years for this mission is long enough to cover an entire solar cycle. Tens of medium-strong events are estimated to be detected by LISA providing for the first time a sampling of SEPs at small steps in longitude with the same detectors avoiding typical normalization problems found among different experiments. The role of solar physics with and for LISA was described in detail in [27].

ASTROD offers analogous but complementary possibilities compared to LISA in case radiation monitors would be placed on board. Analogous because, in order to monitor the test-mass charging, is plausible that radiation monitors on board ASTROD will be optimized for particle energy measurement above 100 MeV(/n) but the different orbits of the ASTROD spacecraft with respect to those of the LISA missions will provide complementary observations of solar events.

The ASTROD I spacecraft will move from a few hundreds of km from Earth to a maximum distance of 1.7 AU. ASTROD II consisting of a constellation of three spacecraft, with one placed near Earth in the Lagrangian point L1 and the other two moving between 1 and 2 AU from Earth, will allow for observation of solar events magnetically well connected to Earth as well as of events generated in the far side of the Sun. To this purpose, it is still an open discussion among people working on space weather about the necessity of a far-side solar observatory for solar energetic particle risk mitigation strategy. More possibilities would be offered by ASTROD-GW and Super-ASTROD.

We point out that measurements carried out by radiation monitors on board space interferometers devoted to gravitational wave detection might be correlated to those of solar physics experiments. In particular STEREO [28] that, however, will not allow for particle energy measurements above 100 MeV(/n). Other planned multipoint observatories are Solar Sentinels [29] that with Solar Probe [30] and Solar Orbiter [31] are supposed to take data from a few Sun radii to1 AU. However, Solar Probe Plus is designed for protons and nucleus detection up to 100 MeV/n. Therefore the ASTROD I measurements are complementary of those that will be carried out on Solar Probe. The particle package on Solar Orbiter was designed for proton and ion detection up to 400 MeV and 100 MeV/n, respectively. In general, ASTROD I, Solar Probe Plus and Solar Orbiter might not take data at the same time providing a long-term monitoring of high energy solar particles. For the time that these missions would orbit contemporaneously, the resulting multipoint solar event evolution observations would be very interesting despite the typical normalization problems affecting different experiments stressed above.

**5. Discussions and outlook**

5.1 Telescope size

The optical concept includes a broadband high reflective coating (BHRC) for all telescope components that are exposed to direct or indirect sunlight, like the rear side of the secondary mirror, baffles, the telescope tube and last not least the telescope main and secondary mirrors. The pinhole device is also coated with BHRC. The spatial sunlight filtering (pinhole) only allows the entering of sunlight from the target direction into the optical system inside the spacecraft. Therefore most of the sunlight is reflected back to space. Modern coating technologies allow reflectivity values of > 99.8% for narrow band dielectric coatings and 99% for broadband metallic coatings.

Taking the solar constant of 1367 $W/m^2$ at 1 AU, the value becomes 5468 $W/m^2$ at 0.5 AU. This results in a total incoming sunlight power of < 390 W for an aperture of 0.3 m. If 1% of this amount is absorbed by the telescope corresponding to a reflectivity of 99% the residual heat load is only 3.9 W which is not a major problem.

Nevertheless, the telescope system will be connected thermally to the spacecraft radiative panels. The thermal properties and behaviour will have to be addressed in more detail during the further development as well as details of possible surface coating degradation due to high energy particles.

If for some reason the telescope diameter would have to be halved it would have an impact on the S/N ratio of the pulse detection. The ratio of sunlight/laserlight would not be affected, but the number of photons per pulse received by the telescope. The number of photons per pulse received from earth around the opposition (1.7 AU) is 5000. A halved telescope diameter would reduce this number to 1250. In principle this amount of photons should be sufficient for the measurement and laser pulse identification, but of course 5000 photons would leave more margin. The effect on the emitted beam is that the earth laser stations receive less collimated beams, and this is compensated by the larger aperture of the earth telescopes.

To compensate the reduced number of 1250 photons per pulse due to a reduced telescope diameter, the ground station laser power could be increased and maybe also the onboard laser power.

5.2 Drag-free tradeoffs

We require continuous use of the drag-free system. The purpose is to maintain as little deviation as possible to the geodesic orbit in the long run. The tolerance for any short run is 3 ps (0.9 mm). Therefore, within this tolerance, the use of drag-free system could be intermittent. This option should be evaluated seriously in the assessment study.

We have done a simulation with the acceleration downgraded by a factor 3, the impact to science goal is that the accuracy of parameter determination would be downgraded by a factor of about 2.5; e.g., for the Eddington parameter γ, the uncertainty would be $7 \times 10^{-8}$ (downgraded by a factor of 2.3), for nonlinear parameter β, $1.6 \times 10^{-5}$ (downgraded by a factor of 2.5); this would still be a valid mission goal.

What is driving the science results is the performance around 0.1 mHz. The performance at 0.36 mHz could be relaxed to be the same as at 0.1 mHz, i.e., $3 \times 10^{-14}$ m s$^{-2}$ Hz$^{-1/2}$.

5.3 Thermal issues

The relative thermal influx between 0.4 and 0.8 AU varies from 100 % to 400%. There are different measures to mitigate the thermal effects and to minimize the corresponding impacts on the science performance. In order to keep the payload temperature variation due to the thermal influx from the sun below delta T = ± 1K the thermal shield must have a high reflectivity and must be thermally decoupled from the spacecraft as good as possible. Second surface mirrors on the thermal shield improve the efficiency. Waste heat from the spacecraft (including the heat produced by the payload) is transported via heatpipes to the cold surface of a radiator facing to the cold deep space. Additional Multilayer Insulation (MLI) can be used to control the thermal behaviour of the spacecraft and the payload according to the requirements. For an active control of the thermal budget thermal control louvers can be used. Typical thermal control louvers performance data are shown in the figure. This technology is well established since decades for spacecraft thermal control tasks.

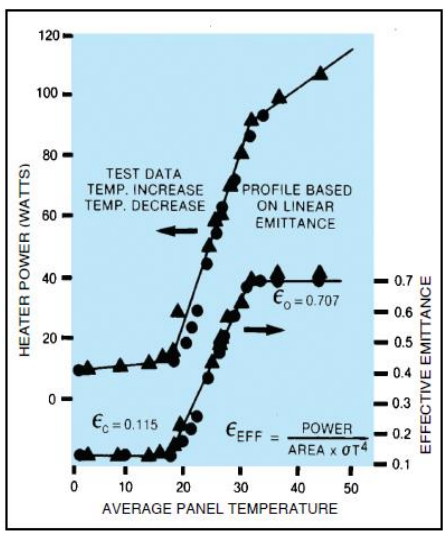

Fig. 11. Typical Louvers performance data (Orbital - Technical Services Division Courtesy)

The measurement principle demands a pointing of the telescope towards the earth during the mission. During the solar oppositions the telescope is exposed to direct sunlight. Fortunately these oppositions coincide with the maximum sun distances of around 0.8 AU. The direct thermal influx around the minimum sun distances (~0.4 AU) can easily be avoided by use of a sunshade system, because here the angle between the telescope axis and the direction to the sun is quite large. The thermal flux into the spacecraft by direct sunlight is minimized by the optical layout of the telescope components. Nevertheless, a residual thermal difference between the solar oppositions with direct influx and the phases without direct influx will remain, resulting in different net heat loads. Possible distortions of the telecope components due to different temperatures will have to be calculated in detail, but with state of the art substrate materials and the additional possibility of optical light path correction by active optical elements there are no unsolvable problems to be expected.

5.4 Outlook

A natural series would consist of ASTROD I and ASTROD-GW [32, 33, 34, 35]. The ASTROD-GW would be dedicated to probe gravitational waves at frequencies below the LISA bandwidth to 100 nHz and to detect and study solar g-modes. ASTROD-GW, with best sensitivity between100 nHz and 1 mHz, will compliment LISA and PTAs (Pulsar Timing Arrays) to study co-evolution of galaxies with black holes, and dark energy. ASTROD-GW will also compliment DECIGO and Big Bang observer to study inflationary gravitational waves. ASTROD-GW assumes similar level of power requirement and strain (acceleration) noise requirement as LISA; but with the strain noise requirement extended to lower frequency to be implemented using absolute laser metrology. The M3 frame cost would be enough for ASTROD I with the development of LISA Pathfinder; significant reduction may come if LISA Pathfinder over-performs in the strain noise budget which has been actually aimed by its team members, or the ASTROD I strain noise requirement reduced by a factor of 3. The cost of ASTROD-GW, after LISA, would be similar to LISA, or less with a number of facilities already established (e.g. deep space networks) before implementation.

The field of the experimental gravitational physics stands to be revolutionized by the advancements in several critical technologies, over the next few years. These technologies include deep space drag-free navigation and interplanetary laser ranging. A combination of these serves as a technology base for ASTROD I. ASTROD I is a solar-system gravity survey mission to test relativistic gravity with an improvement in sensitivity of over 3 orders of magnitude, improving our understanding of gravity and aiding the development of a new quantum gravity theory; to measure key solar system parameters with increased accuracy, advancing solar physics and our knowledge of the solar system; and to measure the time rate of change of the gravitational constant with an order of magnitude improvement and the anomalous Pioneer acceleration, thereby probing dark matter and dark energy gravitationally. This will be the beginning of a series of precise space experiments on the dynamics of gravity. The techniques are becoming mature for such experiments.